# A Simple Method of Demonstration of Characteristics of Rainbows Using a Glass of Water and a Few Laser Sources

Soumen Sarkar[1,4], Sanjoy Kumar Pal[2,4], Pradipta Panchadhyayee[3,4*], Shinjinee Das Gupta[5], and Debapriyo Syam[6]

[1]Karui P.C. High School, Hooghly, West Bengal, India
[2]Anandapur High School, Anandapur, Paschim Medinipur, West Bengal, India
[3]Department of Physics (UG & PG), Prabhat Kumar College, Contai, Purba Medinipur, India
[4]Institute of Astronomy Space and Earth Science, Kolkata -700054, W. B., India
[5]Department of Physics, Victoria Institution (College), Kolkata-700009, W. B., India
[6]Kalpana Chawla Centre for Space and Nanosciences, Kolkata - 700010, India
[*]E-mail: pradipta@pkcollegecontai.ac.in

## Abstract

A rainbow is a captivating natural phenomenon resulting from the refraction, dispersion, and reflection of sunlight within water droplets. Traditional classroom demonstrations often focus on qualitative explanations of the formation of rainbows using prisms or water bowls. This study presents a simple experimental approach to analysing the process of rainbow formation through quantitative analysis using a cylindrical glass filled with water, graph paper, and three semiconductor laser sources emitting red, green, and blue light. By measuring the angles of minimum deviation for different wavelengths, we have found that the experimental values closely match the theoretical predictions. This method offers a hands-on, cost-effective approach to enhance students' understanding of the physics behind rainbows.

## 1. Introduction

A rainbow is one of nature's most spectacular optical phenomena, which is often seen after a rain shower when sunlight interacts with water droplets in the atmosphere. Understanding the physics behind rainbows enhances our appreciation of this natural phenomenon, bridging the gap between science and beauty. Rainbows are formed due to the refraction, dispersion, and reflection of sunlight within water droplets. When light enters a raindrop, it bends due to refraction. Inside the droplet, light is dispersed into its constituent colours and reflected back (once or two times) from the inner surface of the drop before emerging in the generally backward direction. The combination of these effects creates a circular arc of colours in the sky. The colours appear in the order (outer to inner edge) of red, orange, yellow, green, blue, indigo, and violet for the primary rainbow and in the reverse order for the secondary rainbow (see Fig. 1).

Several scientists, including Newton, Young, Airy, and Mie, have contributed towards the explanation of the various aspects of the rainbows [1- 4]. In some papers, mathematical equations and modelling are employed to explain the formation of the rainbows [5,6]. In recent times, Dragia T. Ivanov and Stefan N. Nikolov proposed an alternative method for demonstrating the rainbow. They used a large glass sphere with a reflective coating using common kitchen aluminium foil [7]. Craig Bohren used a beaker filled with water and a laser beam to demonstrate and explain the appearance of rainbows [8]. Qualitative and quantitative laboratory experiments on mirages, rainbows, and halos, incorporating both demonstration and comparison with light-scattering simulations, were reported by M. Vollmer and R. Tammer[9]. The angular positions of rainbows formed by single drops were studied using a student spectrometer, revealing dependence on refractive index and drop size. Experiments with water and corn syrup demonstrated multiple rainbow orders and polarisation effects, which were verified using He-Ne laser interference patterns [10]. Outstanding cylindrical rainbow experiments were discussed by Boyer[11].

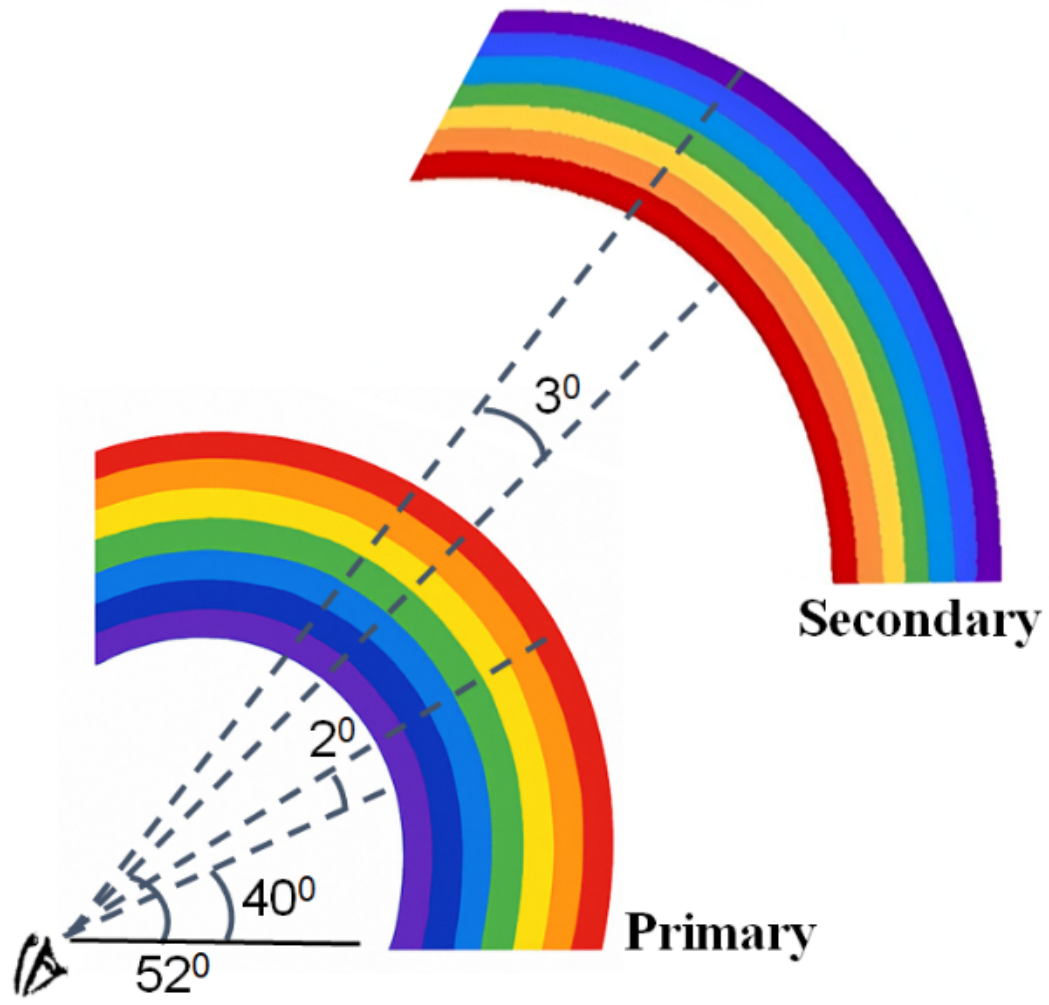

Fig. 1: Arrangement of colours in primary and secondary rainbows.

In various secondary and higher secondary classes, we frequently discuss the issue of rainbow formation from a qualitative viewpoint. Usually, in classrooms, teachers demonstrate the formation of rainbows using a prism or a setup with a plane mirror and a bowl of water. To facilitate a better understanding of the intricacies of optical phenomena among students, this paper presents a simple yet innovative experiment that can be conducted at home or in a school laboratory. Our method requires a cylindrical glass vessel filled with water, graph paper, and three laser sources that emit red, green, and violet light. To verify that the order in which bows of different colours are arranged as seen by an observer matches the order found by our method, we have measured the angles of minimum deviation for red, green and violet light. This experiment is designed for high school students (students in grades XI and XII) who may not have access to spectrometers. It can be implemented as a small-scale project involving a group of students. For safety, a Class 2 or Class 3R laser source (<5 mW) should be used. Additionally, students should be advised to wear protective eyewear, such as laser safety goggles or sunglasses, during the experiment.

## 2. Theory

In explaining rainbow formation, we usually consider the example of polychromatic light passing through a spherical liquid drop. Here, we have used a cylindrical glass tumbler filled with water in our setup, which can be assumed to be an analogue of a drop, as far as propagation in the horizontal plane is concerned. The reason why a cylindrical glass of water or even a glass of water having the shape of a truncated cone exposed to a horizontal ray of light behaves like a spherical bulb of water is explained below with the help of Fig. 2.

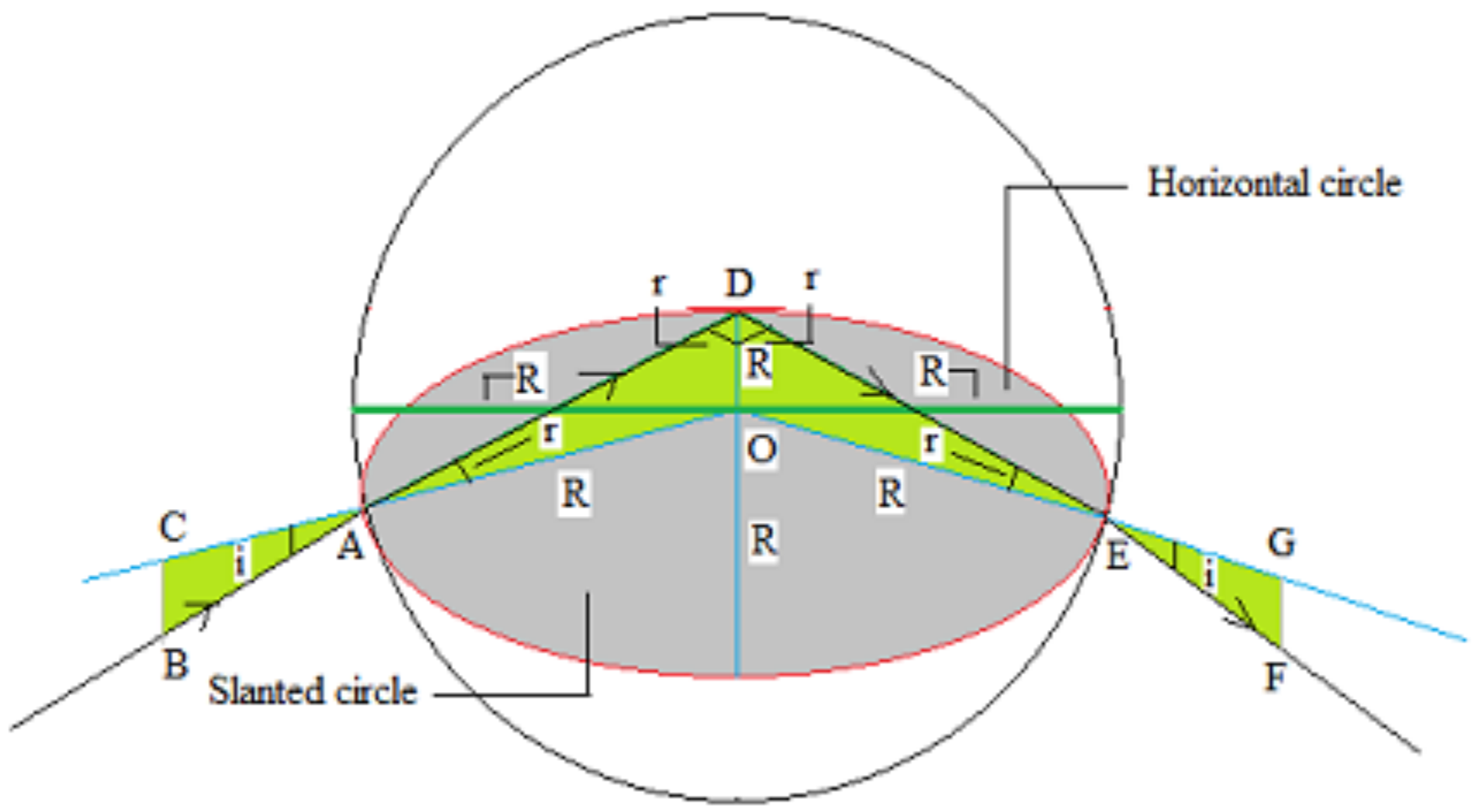

Fig. 2: Schematic diagram showing the propagation of light through a spherical bulb of water.

It shows a ray of light passing through a transparent spherical container, centred at O, filled with water. BA is the incident ray. Note that, by one of the laws of refraction, the plane AOD of refraction coincides with the plane of incidence, ABC. Again, the plane DOE defined by the reflected ray and OE coincides with the plane AOD defined by AD and OA. Finally, the plane EGF of the emergent ray coincides with the plane ODE. Thus, the ray BA, as it propagates through the spherical bulb and emerges as EF, always lies in one plane. Incidentally, that plane cuts the bulb in such a way that the cross section of the plane bounded by the bulb is a circle. Precisely the same thing happens to a horizontal ray of light that falls on a cylindrical glass of water. Having established the logic behind using a cylindrical glass of water instead of a spherical bulb of water, it is imperative to consider all relevant parameters. In this regard, a natural question arises: how does the wall thickness of the container affect the results? A detailed exploration of this aspect is presented in Appendix 1.

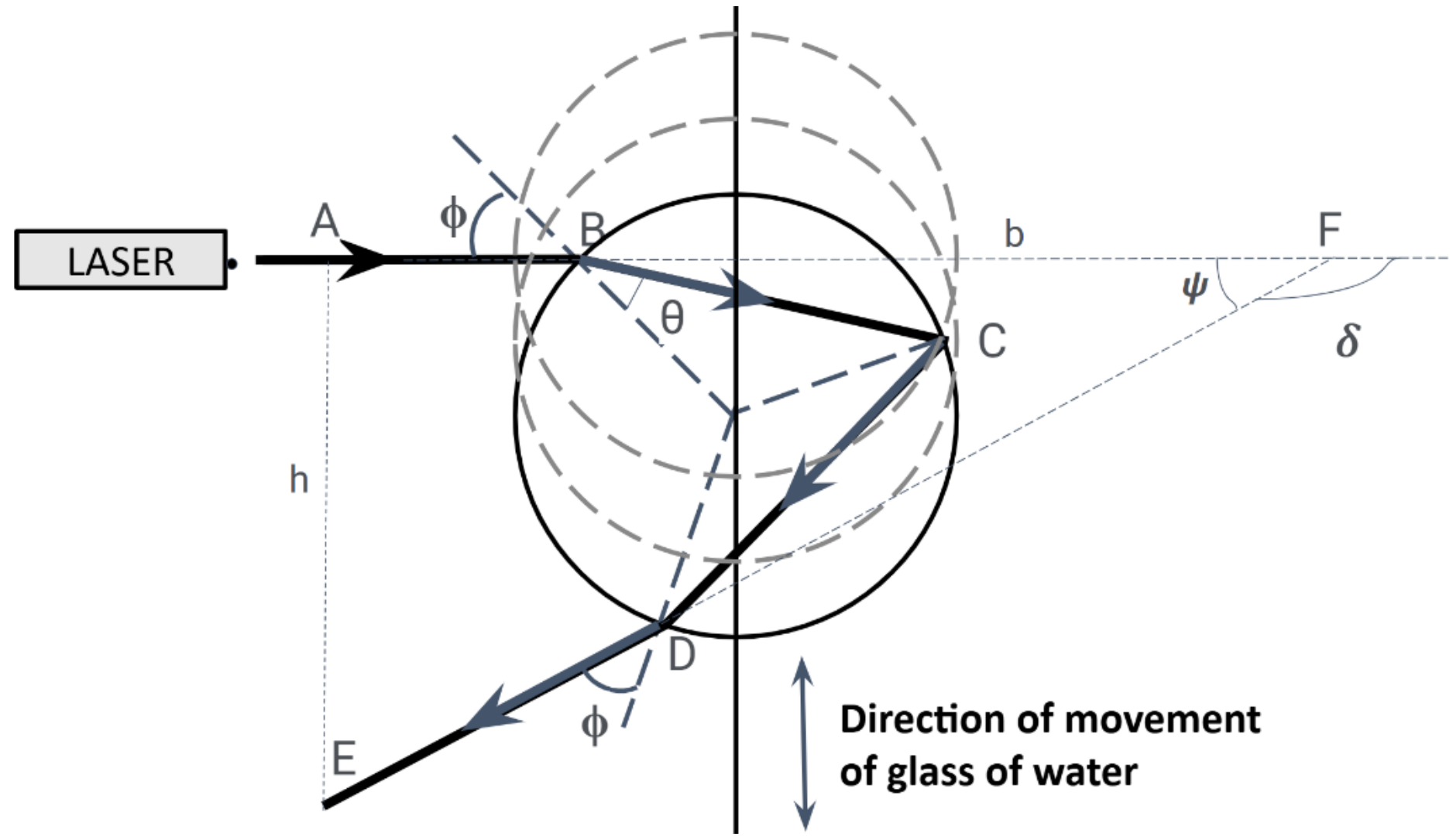

Fig. 3: Schematic diagram of propagation of a monochromatic ray for the formation of a rainbow of the 1st order (primary rainbow).

When a horizontal ray AB enters the water at B, it undergoes partial reflection at C and eventually emerges at D along DE (Fig. 3). The total deviation of the ray is denoted by δ.

$$\delta = (\phi - \theta) + \pi - 2\theta + (\phi - \theta) = 2(\phi - \theta) + (\pi - 2\theta). \quad (1)$$

In the case of two internal reflections at C and D, the total deviation (see Fig.4) is:

$$\delta = 2(\phi - \theta) + 2(\pi - 2\theta), \quad (2)$$

and for $p$ internal reflections:

$$\delta = 2(\phi - \theta) + p(\pi - 2\theta). \quad (3)$$

Since the refractive index of water ($n$) can be expressed as $n = \frac{sin\ \phi}{sin\ \theta}$, we can write,

$$si\,n\,\theta = \frac{\sin\phi}{n}. \quad (4)$$

From Eqs. (3) and (4), it follows that $\delta$ is a function of ϕ, since $n$ remains constant for water in the case of a given wavelength.

To find the value of ϕ for which $\delta$ is minimum, we differentiate $\delta$ with respect to $\phi$ and equate it to zero. So

$$d\delta = 2d\phi - 2d\theta - 2p\ d\theta = 2[d\phi - (p+1)d\theta] = 0.$$

This condition gives, $d\phi = (p+1)d\theta$. (5)

Again, differentiating Eq. (4) we have, $co\,s\,\phi\ d\phi = nco\,s\,\theta\ d\theta$. (6)

Substituting the expression of $d\theta$ (from Eq. (6)) in Eq. (5), we obtain,

$$d\phi = (p+1)\frac{cos\ \phi\ d\phi}{n\ cos\ \theta}.$$

Squaring both sides of the above equation, the following equation is obtained:

$$n^2(1 - sin^2\theta) = (p^2 + 2p + 1)cos^2\phi.$$

So, $n^2 - sin^2\phi = cos^2\phi + (p^2 + 2p)cos^2\phi$.

$$\therefore cos\phi = \sqrt{\frac{n^2-1}{p(p+2)}}. \quad (7)$$

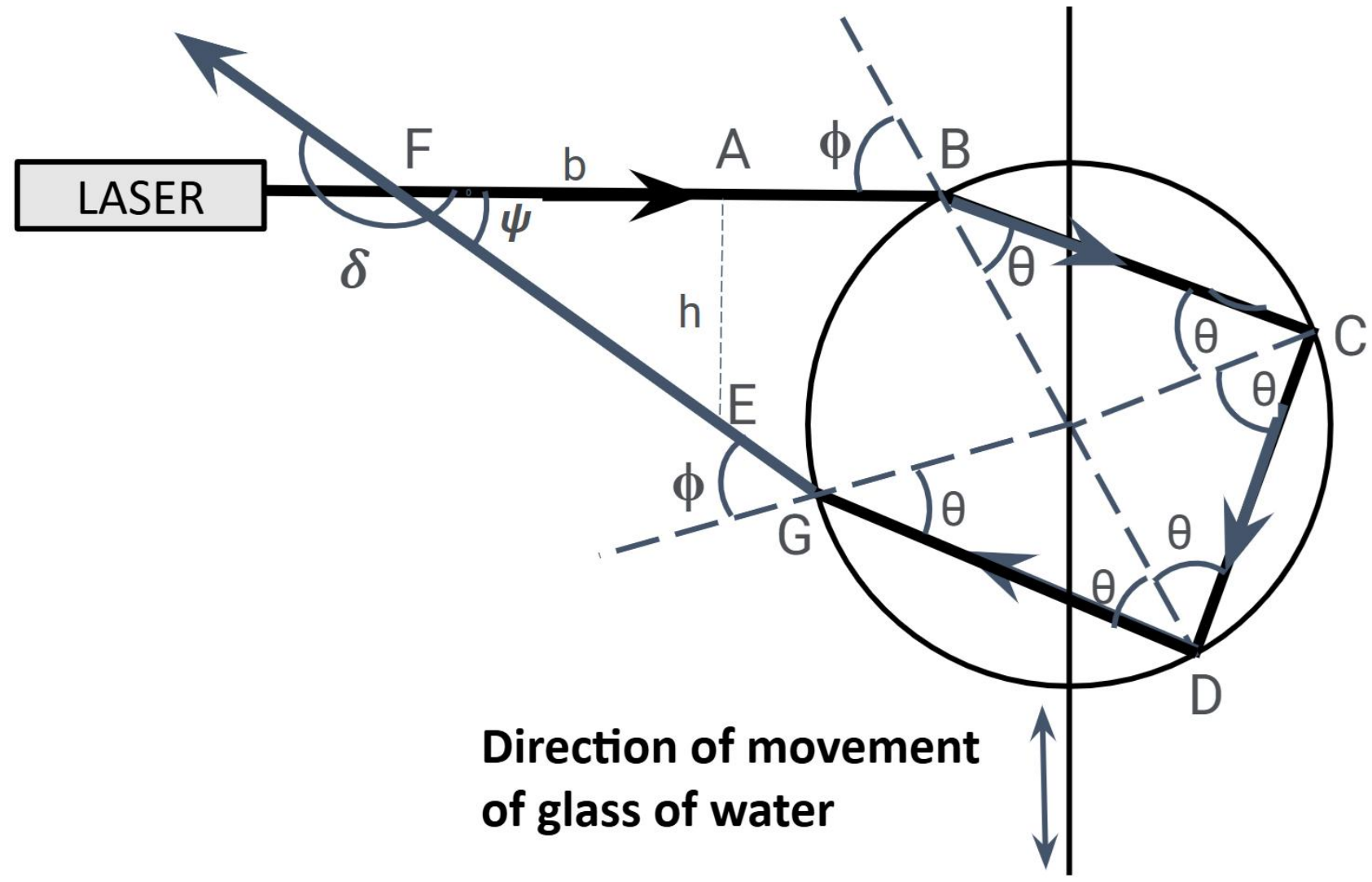

Fig. 4: Schematic diagram of ray propagation for the rainbow of the 2nd-order (secondary rainbow).

For the primary rainbow (1st order i.e. $p = 1$), $\phi = \cos^{-1}\sqrt{\frac{n^2-1}{3}}$. (8)

For the secondary rainbow (2nd order i.e. $p = 2$), $\phi = \cos^{-1}\sqrt{\frac{n^2-1}{8}}$. (9)

To know whether this value of $\phi$ gives a maximum or minimum for $\delta$, we calculate $\frac{d^2\delta}{d\phi^2}$ , and substitute the above value of $\phi$ for minimum deviation.

The results show that $\frac{d^2\delta}{d\phi^2} > 0$.

Therefore, the deduced condition (Eq. (7)) for $\phi$corresponds to the minimum value of $\delta$.

From Eq. (4) and Eq. (7), we can write

$$sin\theta = \frac{\sqrt{1-cos^2\phi}}{n} = \sqrt{\frac{p^2+2p-n^2+1}{n^2(p^2+2p)}} \quad (10)$$

The values of $\phi$ and $\theta$ from Eq. (7) and Eq. (10), respectively, are then substituted in Eq. (3). As a result, we obtain the general expression of the angle of minimum deviation as

$$\delta_{min} = 2\left(cos^{-1}\sqrt{\frac{n^2-1}{p^2+2p}} - sin^{-1}\sqrt{\frac{p^2+2p-n^2+1}{n^2(p^2+2p)}}\right) + p\left(\pi - 2sin^{-1}\sqrt{\frac{p^2+2p-n^2+1}{n^2(p^2+2p)}}\right) \quad (11)$$

For the primary rainbow (1st order), the minimum deviation is

$$\delta_{min} = 2\left(cos^{-1}\sqrt{\frac{n^2-1}{3}} - sin^{-1}\sqrt{\frac{4-n^2}{3n^2}}\right) + \left(\pi - 2sin^{-1}\sqrt{\frac{4-n^2}{3n^2}}\right), \quad (12)$$

and for the secondary rainbow (2nd order), the minimum deviation becomes

$$\delta_{min} = 2\left(cos^{-1}\sqrt{\frac{n^2-1}{8}} - sin^{-1}\sqrt{\frac{9-n^2}{8n^2}}\right) + 2\left(\pi - 2sin^{-1}\sqrt{\frac{9-n^2}{8n^2}}\right). \quad (13)$$

Since the value of $\phi$ for which the deviation is minimum is a function of the refractive index, the angle of minimum deviation varies with the colour of light. The value of $\psi$ (see Fig. 3), for the 1st order rainbow, it can be determined easily from the relation, $\psi = \pi - \delta_{min}$. For the 2nd order rainbow, $\psi = \delta_{min} - \pi$. (See Fig. 4).

## 3. Experimental Details

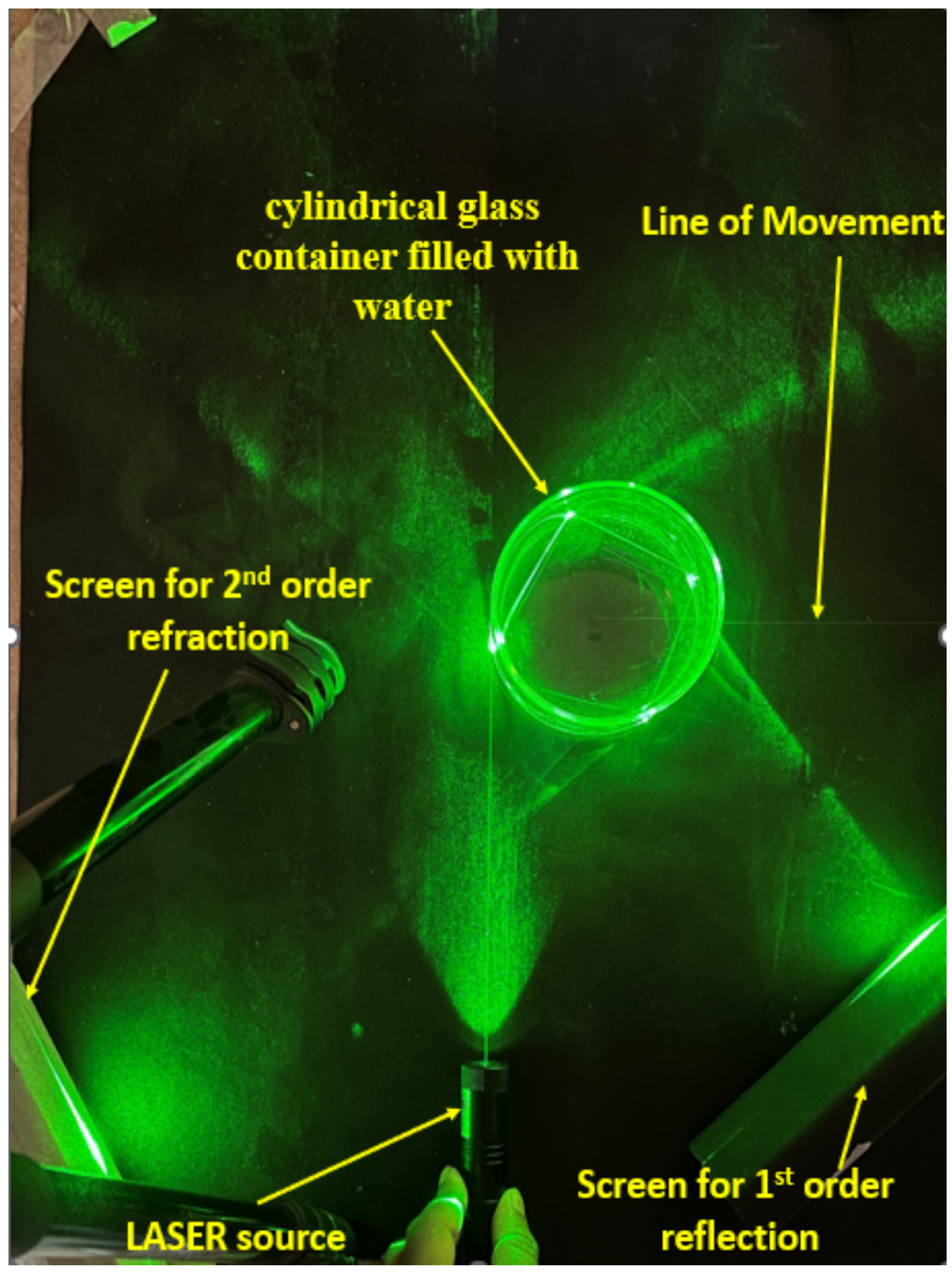

Fig. 5: Setup for analysing laser beam refraction and minimum deviation in a water-filled cylindrical glass (Photo captured without graph paper for clearer visibility of the laser beam).

At the outset, we placed a large millimetre graph paper (50 cm × 50 cm) on a horizontal plane to ensure a stable and level surface for accurate measurements. We have securely fixed laser pointers along the X-axis of the graph paper. We have positioned a cylindrical tumbler (made of glass and filled with tap water) on the graph paper so that the laser beam falls near one end of the diameter of the vessel. The tumbler (outer diameter is 9.6 cm) is so chosen that its wall thickness (1 mm) is thin to avoid (actually minimise) any extra-refraction of the laser beam. As the light enters the water-filled glass tumbler, refraction and reflection occur making the emergent ray visible on the screen. Horizontality of each laser beam was checked at various points by measuring, with a standard ruler scale, the height of the beam above the graph paper. We have then carefully moved the glass along the Y-axis very slowly to find the position where the emergent ray experiences minimum deviation [13]. At this exact point, we have traced the paths of the incident and emergent rays on the graph paper for further analysis. To accurately trace the incident ray, we have nailed two pins onto the graph paper—one near the laser source and another near the point where a spot of light appears on the outer surface of the vessel. Similarly, we have traced the path of the emergent ray using the same method. After marking these paths, we have extrapolated the two rays on the graph paper and constructed a right-angled triangle based on the traced rays. We have measured the height and the length of the base of the triangle using the rulings of the millimetre graph paper directly to compute the angle of minimum deviation ($\delta$). Thus, we have determined the corresponding supplementary angle ($\psi$). We have repeated this entire procedure using three lasers emitting different colours (red, green and violet) of light. We have experimented with the primary (1st order) as well as the secondary (2nd order) rainbows, as shown in Fig. 5.

## 4. Results and Discussions:

Here, we have used three wavelengths of light emitted by three different semiconductor laser sources (Class 3R: output power <5 mW), namely 405 nm (violet), 532nm (green) and 635nm (red), as mentioned by the manufacturer (Climberty) in the specification data pasted on the laser sources. The reasonable values of refractive indices of water at room temperature (25°C) are 1.343± 0.001, 1.339 ± 0.004, and 1.332 ± 0.001, respectively, for those three colours [14]. From Table 1, we observe that the experimental values of $\psi$are nearly equal to the theoretical ones.

Table 1: Comparison of theoretical and experimental results

| Order | Colour | Wavelength (nm) | Refractive index, $n$ | Height, $h$ =AE (cm) | Base, $b$ =AF (cm) | Experimental value of $\psi$= $\tan^{-1}(h/b)$ (deg) | Theoretical value of $\psi$ (deg) |
|---|---|---|---|---|---|---|---|
| First (Primary) | Violet | 405 | 1.343 | 31.3 | 37.0 | 40.2 ± 0.05 | 40.7 |
| | Green | 532 | 1.339 | 32.8 | 37.2 | 41.4 ± 0.06 | 41.2 |
| | Red | 635 | 1.332 | 33.1 | 36.3 | 42.4 ± 0.05 | 42.3 |
| Second (Secondary) | Violet | 405 | 1.343 | 7.9 | 5.8 | 53.7 ± 0.26 | 53.4 |
| | Green | 532 | 1.339 | 7.3 | 5.6 | 52.5 ± 0.28 | 52.4 |
| | Red | 635 | 1.332 | 6.3 | 5.2 | 50.5 ± 0.31 | 50.6 |

[Calculations of the standard deviation of $\psi$ are given in Appendix 2]

## 5. Conclusion:

This study successfully demonstrates a quantitative and straightforward approach to understanding rainbow formation using a cylindrical water container and laser sources of different wavelengths. After measuring the angles of minimum deviation for the primary and secondary rainbows, we have compared these values with the theoretical ones. The experimental results are in close conformity with theoretical predictions. The deviations (though relatively small) of experimental results are primarily due to measurement uncertainties for the graph paper and refraction at the wall of the tumbler. This method provides a cost-effective and accessible way for students to explore the physics behind rainbows, offering a quantitative alternative to conventional qualitative demonstrations.

**Appendix 1**

**Effect of Container Wall Thickness on Experimental Results:**

This issue arises irrespective of whether the bulb contains water or is evacuated. For simplicity, we consider an evacuated bulb. (The relevant diagram is given on the next page.) O is the centre of the bulb, R its radius, and t its thickness. Dashed lines represent the path of the ray when the thickness of the glass is negligible (This is the line APDFHI). When the thickness (t) of glass is small but non-negligible, the path of the ray is ABC up to C. Then it splits into two rays. Part of the incident ray is reflected from the inner surface of the bulb and proceeds along CEJK. The other part of the ray that is incident at C goes to D, where it is reflected from the outer surface of the glass bulb. It then proceeds along DGHI and emerges as if the glass has no thickness. There are additional reflected and refracted rays at A, B, D, E and G, which are not shown for clarity. The lateral separation between the two emergent rays is calculated below.

Referring to the diagram, in the simplest approximation,

$$PB = (\tan i - \tan r)\, t = d\ (say)$$

BC is parallel to the incident ray.

$$QC = PB = d$$

The separation between the rays DF (dashed line) and the ray SG is also the same (d).

CE = DF.

Hence

$$2t.\tan r + QC = 2t.\tan r + (\tan i - \tan r)\, t = (\tan i + \tan r)t$$

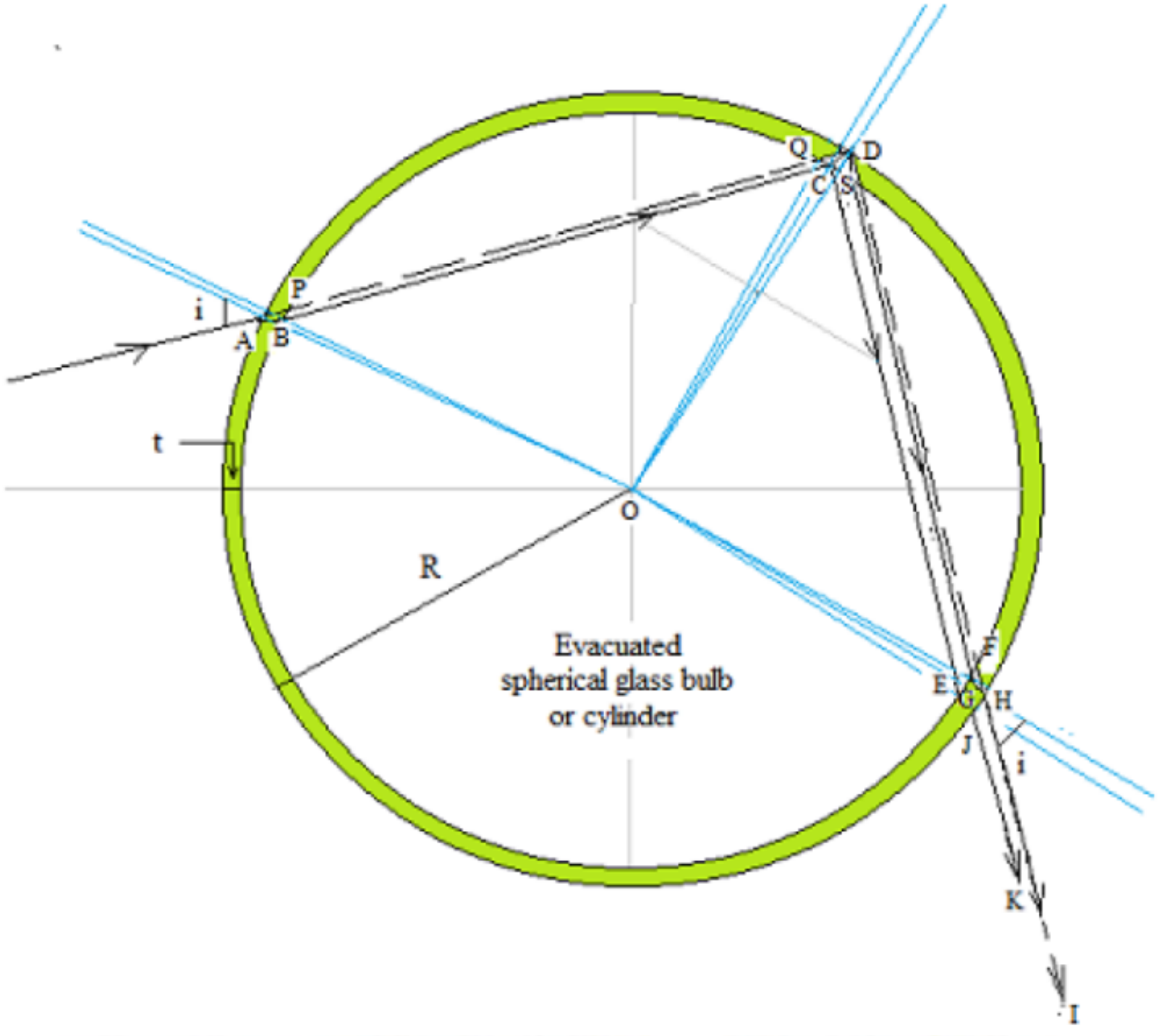

Fig. 6: Schematic diagram showing the effect of the thickness of the wall of the container

**Finally, the perpendicular distance between the rays JK and HI is** 2. tan r.**cos *i*.**

If $i$=45° and $\mu_{glass} \equiv \mu$=1.5,

$$\sin r = \frac{\sin i}{\mu} = 28.43^0$$

Therefore

$$\tan r = 0.5346$$

$$2.t.\tan r.\mathbf{cos}\,\boldsymbol{i} = 2 * 0.1 * 0.5346 * 0.707 = 0.0756\,cm$$

The width of the LASER beam is ~ 0.1 cm. **Thus, the width of the LASER beam is a little more than or about the same as the distance between the rays JK and HI; glass thickness does not matter.**

In a higher order of approximation:
The ray that emerges into the cavity at P makes an angle ‘j’ with OP.

$$\sin j = \sin(i+\eta) \cong \sin i + \cos i \,.\, \eta$$

Again

$$\sin j = \mu \sin (r+\varepsilon) \cong \mu \sin r + \mu \cos r \,.\, \varepsilon$$

Thus

$$\eta \cong \frac{\mu \cos r \,.\, \varepsilon}{\cos i}$$

Now

$$BP \cong t.\tan r$$

Also

$$\varepsilon \cong \frac{BP}{R} = \frac{t.\tan r}{R}$$

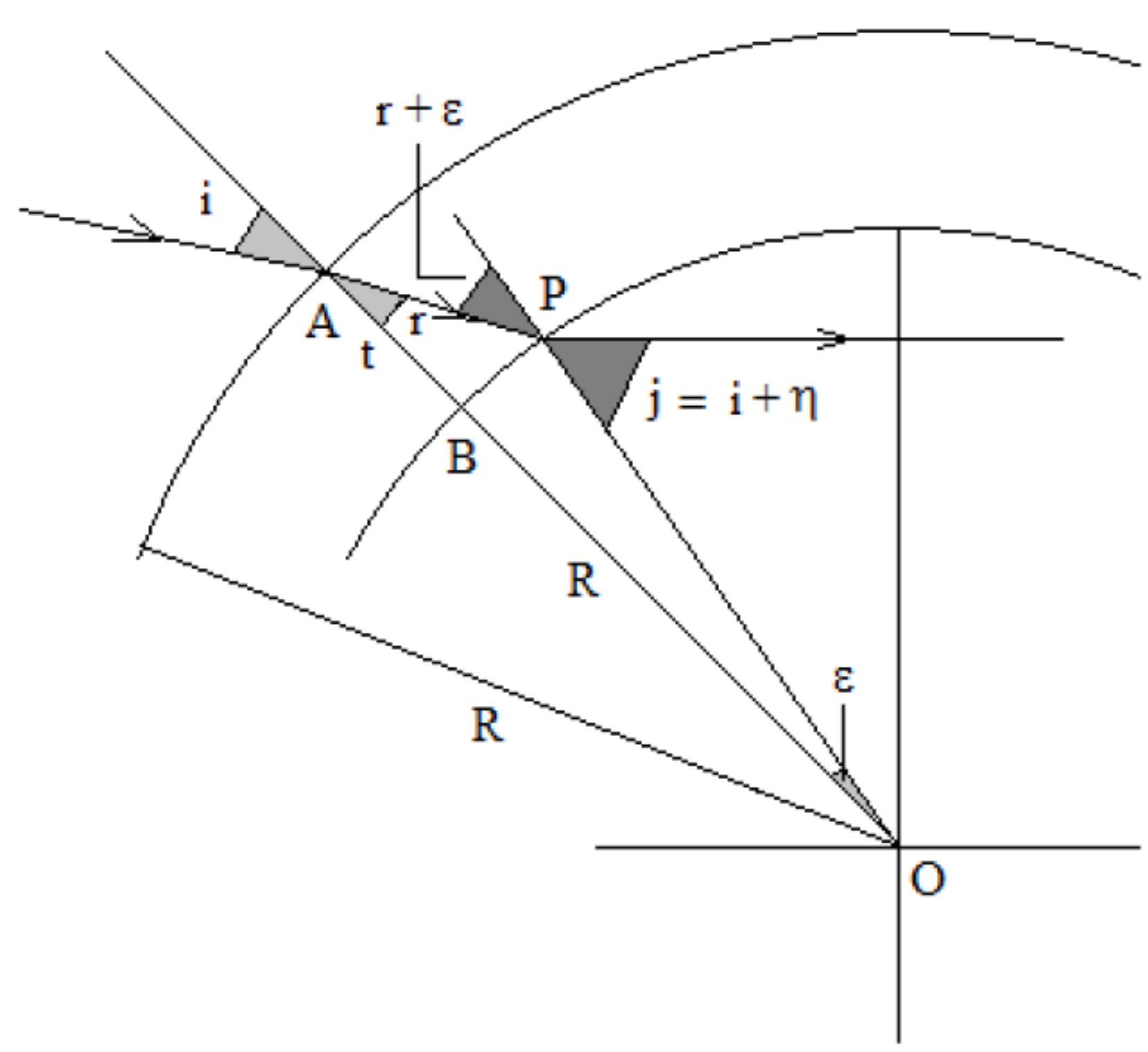

Fig. 7: Schematic diagram showing the effect of the thickness of the wall of the container in higher order approximation

So, finally

$$\eta \cong \frac{\mu \cos r \,.\, \varepsilon}{\cos i} \cong \frac{\mu \cos r}{\cos i} \cdot \frac{t.\tan r}{R} \cong \frac{t.\tan i}{R}$$

When

$$t = 0.1 \text{ cm}, \qquad R = 4.8 \text{ cm}, \qquad \tan i = \tan 45^\circ \;= 1\; \eta \cong 0.021\, rad = 1.194^\circ.$$

Which is a small number. In any case, due to the spherical symmetry of the glass bulb or the axial symmetry of the cylindrical glass vessel, the separation between the two emergent rays (JK and HI) would still be about 0.1 cm.

**Appendix 2**

**Error estimation:**

We are required to estimate the standard deviation of $tan^{-1}\left(\frac{h}{b}\right)$.

Recall that

$$d(tan^{-1}x) = \frac{dx}{1+x^2}$$

Let $\delta x$ denote the standard deviation of x. Then, noting that $b$ and $h$ are independent variables,

$$\delta\left(tan^{-1}\left(\frac{h}{b}\right)\right) = \frac{b^2}{h^2+b^2}\cdot\sqrt{\left(\frac{1}{b^2}(\delta h)^2\right)+\left(\frac{h^2}{b^4}(\delta b)^2\right)}\cdot$$

$$= \frac{1}{h^2+b^2}\cdot\sqrt{((b\delta h)^2)+((h\delta b)^2)}\ \text{(radians)}$$

*Calculations:*

For $\delta h = \delta b = 0.1$ cm (smallest division of graph paper), for 5 repeated observations, the standard deviation <u>in the mean</u> becomes

$$\delta\left(\tan^{-1}\left(\frac{h}{b}\right)\right) = \frac{1}{\sqrt{5}}\frac{0.1}{\sqrt{h^2+b^2}}$$

For $h$ = 31.3 cm, $b$ = 37.0 cm, $\delta\left(\tan^{-1}\left(\frac{h}{b}\right)\right) = \frac{0.00206}{2.236}$ rad ≅ 0.05 °

$h$ = 32.8 cm, $b$= 37.2 cm → $\delta\left(\tan^{-1}\left(\frac{h}{b}\right)\right) = \frac{0.002201}{2.236}$ rad ≅ 0.06°

$h$ = 33.1cm, $b$ = 36.3 cm → $\delta\left(\tan^{-1}\left(\frac{h}{b}\right)\right) = \frac{0.00204}{2.236}$ rad ≅ 0.05°

Again, for $h$ = 7.9 cm, $b$ = 5.8 cm → $\delta\left(\tan^{-1}\left(\frac{h}{b}\right)\right) = \frac{0.0102}{2.236}$ rad ≅ 0.26°

$h$ = 7.3 cm, $b$ = 5.6 cm → $\delta\left(\tan^{-1}\left(\frac{h}{b}\right)\right) = \frac{0.0108}{2.236}$ rad ≅ 0.28°

$h$ = 6.3 cm, $b$ = 5.2 cm → $\delta\left(\tan^{-1}\left(\frac{h}{b}\right)\right) = \frac{0.0122}{2.236}$ rad ≅0.31°